\documentclass[num-refs]{wiley-article}

\usepackage{siunitx}

\papertype{Chapter}

\title{The Life and Death of Software Ecosystems}
\author[1\authfn{1}]{Raula Gaikovina Kula}
\author[2\authfn{2}]{Gregorio Robles}

\affil[1]{Raula is the main contributor of attractors and detractors (i.e., life) to FLOSS Projects}
\affil[2]{Gregorio is the main contributor for the death of ecosystems}

\corraddress{\authfn{1} Assistant Prof. Raula Gaikovina Kula, Nara Institute of Science and Technology, Japan\\ email:raula-k@is.naist.jp\\
\authfn{2} Associate Prof. Gregorio Robles, Universidad Rey Juan Carlos, Spain\\ email:grex@gsyc.urjc.es}

\begin{document}

\maketitle

\begin{abstract}
Software ecosystems have gained a lot of attention in recent times. Industry and developers gather around technologies and collaborate to their advancement; when the boundaries of such an effort go beyond certain amount of projects, we are witnessing the appearance of Free/Libre and Open Source Software (FLOSS) ecosystems.

In this chapter, we explore two aspects that contribute to a \textit{healthy} ecosystem, related to the attraction (and detraction) and the death of ecosystems.
To function and survive, ecosystems need  to attract people, get them on-boarded and retain them. 
In Section One we explore possibilities with provocative research questions for attracting and detracting contributors (and users): the lifeblood of FLOSS ecosystems.
Then in the Section Two, we focus on the death of systems, exploring some presumed to be dead systems and their state in the afterlife.
\end{abstract}








\section{Attractors (and Detractors) to FLOSS Projects}
A contributing component to the sustainability (i.e., life) of a FLOSS project is its ability to attract new development. 
Although keeping current contributors is equally important, projects risk failure if they are unable to attract a healthy amount of new developers to provide rejuvenation and aid in project evolution, especially in response to ever-changing external forces (i.e., impactful events, new technologies, vulnerabilities and rivals) that affect FLOSS projects.
In this section, we discuss (1) the different forces of attraction (and detraction) that influence contributors to participate in specific projects, (2) the effect of these forces at the ecosystem level, and finally present (3) three provocative research questions to further our understanding of attracting contributors to a project. 

\subsection{Forces of Attraction (and Detraction)}
We classify known forces of attraction as either motivation-related, environmental, or a combination of the two.
Internal project-driven campaigns usually revolve around marketing strategies to attract developers.  
A study by \citet{Storey17} showed that communities of FLOSS projects are shaped through social and communication channels (sometimes referred to as social coding).
Recently \citet{Aniche:2018} confirmed that news channels also play an important role in shaping and sharing knowledge among developers.
Hence, owners of projects could boost their social presence through participation on recent topics from news aggregators such as \texttt{reddit}\footnote{\url{https://www.reddit.com}}, \texttt{Hacker News}\footnote{\url{https://news.ycombinator.com}} and \texttt{slashdot}\footnote{\url{https://slashdot.org}}.
For instance, a project may employ new or well-known or recognizable trademarks that are trending in the news.
Social media outlets and other communication channels can be leveraged to improve project attractiveness (i.e., innovative posts on Q\&A forums such as \texttt{StackOverflow}\footnote{\url{https://stackoverflow.com}} and social media endorsements and collaborations through \texttt{twitter} or \texttt{facebook}).
Recently, analytical indicators of project health or fitness are aimed at increasing the appeal of a project.
In detail, the emergence of online collaboration platforms \texttt{GitHub}, \texttt{GitLab} and \texttt{BitBucket}, with specific features such as pull requests, forks, and stars depicts the fitness of a project.

Other motivations are driven by external forces.
\citet{HataChase15} used game theory to identify three strategies that is likely to incite contributions.
The authors suggest that improving the code writing mechanisms (i.e., wikis, offical webpage, contributing and coding guidelines and using multi-language formats).
Secondly, in terms of monetary incentives, sites such as bountysource website\footnote{\url{https://www.bountysource.com/}} allow developers to be hired as bounty hunters to fix specialized bugs in a project. 
Finally, the impact of innovations such as social coding, introduced by online collaborations of \texttt{GitHub} has attracted attention of developers.
A lesser explicit form of motivations is driven by a third-party with their own interests. 
For instance, a company may allocate employees or provide monetary incentives to support (i.e., keep alive) a project of interest.
This is especially in cases where a third-party is interested in stimulating further feature development of an existing product that they are invested in.


Failing projects provide insights into some environmental forces that detract developers from making contributions.
A study by \citet{Coelho:FSE2017} found the following reasons for failing projects: usurped by competitor, obsolete project, lack of time and interest,  outdated technologies, low maintainability, conflicts among developers, legal problems, and acquisition. 
To mitigate these detractors, the authors propose three strategies to rejuvenate contributions in failing FLOSS projects.
Firstly, projects are encouraged to improve their stability by \textit{moving towards an organization account instead of a personal account.}
Secondly, failing projects are encouraged to \textit{transfer the project to new maintainers.}
This is especially needed if the current maintainers' activity has been deteriorating over time.
Finally, the project is encouraged to \textit{accept new core developers}.
This organizational factor aims to rejuvenate and ignite fresh ideas, giving new life to the project.

\subsection{Forces at the Ecosystem Level}
To date, existing works performed their analysis in respect to individual projects. 
At a higher level of abstraction, there exists cases where the forces of attraction (and detractions) in several projects in an ecosystem are triggered by a common event.
For instance, several studies  \citep{Abdalkareem:2017, Decan:2017, Kikas:2017} investigated the eventful case of the JavaScript \texttt{"left-pad"} incident (see~\cite{Web:left-pad}), where removal of a trivial library package caused major breakages in thousands of projects including notable JavaScript frameworks like babel and react.

Other examples of impactful events at the ecosystem level include responses to wide-spreading high risk security vulnerabilities (i.e., \texttt{ShellShock}, \texttt{Heartbleed} and \texttt{Poodle}), rivaling technologies (i.e., battles between competing frameworks for specific programming domains such as PHP\footnote{A blog post for 2018 best PHP frameworks at \url{https://coderseye.com/best-php-frameworks-for-web-developers/}} and JavaScript\footnote{A blog that shows the trend changes between rival JavaScript frameworks\url{https://stackoverflow.blog/2018/01/11/brutal-lifecycle-javascript-frameworks/}}) and
inadequacies in the current situation. 
As an example, current inadequacies could be realized when a change in management occurs (i.e., such as change of the middle-man in InnerSource\footnote{InnerSource takes the lessons learned from developing FLOSS and applies them to the way companies develop software internally. Taken from \url{https://paypal.github.io/InnerSourceCommons/}}). 
Changes in management (i.e., especially the single movement of a key contributor) may set off a chain series of attract and distraction forces that leave behind a rippling effect across the ecosystem. 
We theorize that these forces impact ecosystem sustainability, especially if affected projects act as hubs within that ecosystem. 

\subsection{Provocative Research Questions}
To conclude this section, we formulate a set of provocative research questions to further our understanding of attraction and detraction forces:

\begin{itemize}
\item \textbf{What are the strengths and successes of known attractor strategies to FLOSS projects?}
We have identified many attraction forces. 
Understanding the strength and success of these different attractors will assist us to treat projects that may be suffering with attracting new contributors to their projects.

\item \textbf{How often are these attractor strategies practiced in the real-world and in respect to different ecosystems?}
It is unknown to what extent and the frequency by which these strategies are practiced by practitioners in recent times.
Furthermore, we are unclear of the environmental and ecosystem conditions required to sustain these attraction forces.

\item \textbf{What are the implications and impact of these forces of attraction at the ecosystem level?}
We theorize that attraction forces may impact the overall ecosystem.
However, it is unclear the extent by which these forces of attraction may affect the sustainability of the overall ecosystem itself.
\end{itemize}

\section{On the Death of Ecosystems}
Software ecosystems have gained a lot of attention in recent times. Industry and developers gather around technologies and collaborate to their advancement; when the boundaries of such efforts go beyond certain amount of projects, we are witnessing the appearance of a software ecosystem. Software ecosystems are complex in nature, as many stakeholders are involved. There are for sure key people (e.g., Guido van Rossum in Python) and projects (such as MySQL in the MySQL ecosystem), but activity follows a decentralized pattern, more in the fashion of stigmergic process as known for instance from colonies of ants.

In this section we want to focus on the death of software ecosystems. While it is known that many FLOSS projects are discontinued, to the knowledge of the authors we haven't found any research on the topic on software ecosystems. We define as the death of a project as having no activity in it for a long period as done in other research works. So, a dead software ecosystems have would have no activity.
It should be noted that other definitions of death could be proposed. One may think of having no users, a loss of interest in the software industry, a decrease in developers and developer interest, etc.

On the other hand, we are not looking after projects, which are defined (i.e., they have a goal) and concrete software solution that has an organizational and logistic structure (a known website, repository, mailing list, etc.). Software ecosystems are built of many projects, which co-ordinate themselves (or not) but that have a relationship that is in general technological (although other types of ecosystems such as the (entire) Apache ecosystem orchestrates around collaboration).   

\subsection{Research Questions} 
Current research literature has so far focused mainly in successful FLOSS systems, to see how they are articulated and organized, in order to derive lessons learned out of these. Our method will be exploratory and based on case studies.
Specifically, we want to address following RQs:

\begin{itemize}
\item \textbf{\textbf{RQ$_1$.} What do we know of dead ecosystems?}\\
We want to approach our study based on real cases of ecosystems that were so in the past, but that are now inactive. So, as a first step, we performed an unstructured search for dead ecosystems, by asking participants in the workshop and then by looking in the web (mainly in the webpages of its projects and in Wikipedia) for more information. The output of this research question is a list of dead ecosystems on which the subsequent RQs will be addressed.

\item \textbf{\textbf{RQ$_2$.} Why are these ecosystems dead?}\\
Once we have identified dead ecosystems in RQ1, we would like to dig into the reasons why these have become inactive. In this regard, we would like to see if the cause of the inactivity can be technology (e.g., becoming an outdated technology), economic (e.g., failure of funding), legal (i.e., patent or license issues), among others. As input of information we will use Google searches on the Internet.

\item \textbf{\textbf{RQ$_3$.} What can we learn from dead ecosystems?}\\
Once we have identified dead ecosystems (RQ1) and have further information into what causes are behind its death (RQ2), our goal is to see if we can extract major insight into the topic. The final goal is, of course, to help software ecosystems to stay "healthy".

\end{itemize}

\subsection{Findings}
Based on research questions in the prior section, this section we discuss and present the findings of each research question. 

\subsubsection{\textit{RQ$_1$} What do we know of dead ecosystems?}

\begin{table}[t]
\begin{center}
\caption{ Summary of the studied dead ecosystems}
\begin{tabular}{l|c|r} \hline
System Name &\multicolumn{1}{c|}{Brief Description} 
& \multicolumn{1}{c|}{Discontinued Date} \\ \hline \hline
Concurrent Versioning System (CVS) & version control &	May, 2008	\\ \hline
FireFoxOS & mobile operating system &	Dec, 2015	\\ \hline
Apache Geronimo & application server &	May, 2013	\\ \hline
Maemo & mobile development platform &	Feb, 2010	\\ \hline
\end{tabular}    \label{tab:RQ1}
\end{center}
\end{table}

During the seminar in Shonan, participants were asked informally regarding open ecosystems that have been discontinued. 
After much discussion, as shown in Table \ref{tab:RQ1}, the following dead systems arose from the discussions.

\paragraph{Control Versioning System (CVS)}

CVS is a version control system, an important component of Source Configuration Management (SCM)\footnote{ website is at \url{https://www.nongnu.org/cvs/}}. 
Using it, you can record the history of sources files, and documents. 
The last version of CVS was published in 2008 (see \url{http://savannah.nongnu.org/news/?group=cvs}).

\paragraph{FirefoxOS}

Firefox OS was a mobile operating system, based on HTML5 and the Linux kernel, available for several platforms. 
It was developed by Mozilla Corporation under the support of other companies and a large community of volunteers from around the world. 
The operating system was designed to allow HTML5 applications to communicate directly with device hardware using JavaScript and Open Web APIs\footnote{ although an official website is not found, the blog of one of the key engineers is an example of its existence \url{https://medium.com/@bfrancis/the-story-of-firefox-os-cb5bf796e8fb} }.

In December 2015, Mozilla announced it would stop development of new Firefox OS smartphones, and in September 2016 announced the end of development.

\paragraph{Apache Geronimo}

Apache Geronimo\footnote{website available at \url{http://geronimo.apache.org/}} is a FLOSS application server developed by the Apache Software Foundation and distributed under the Apache license. 
IBM announced May 14, 2013 that it would withdraw and discontinue support of Apache Geronimo (see \url{http://www-01.ibm.com/common/ssi/rep_ca/1/897/ENUS913-081/ENUS913-081.PDF}).
This was also communicated through their website and mailing lists.

\paragraph{Maemo}

Maemo \footnote{website available at \url{http://maemo.org/intro/}} is a development platform for handheld devices based on debian GNU / Linux. Maemo is mostly based on open-source code and has been developed by Maemo Devices within Nokia in collaboration with many FLOSS projects such as the Linux kernel, Debian, and GNOME.

At the Mobile World Congress 2010, Intel and Nokia announced that they would unite their Linux-based platforms into a single product, called MeeGo. The Linux Foundation canceled MeeGo in September 2011 in favor of Tizen. An emerging Finnish company, Jolla, took Mer, a successor based on the MeeGo community, and created a new operating system: Sailfish OS, and launched a new smartphone at the end of 2013.

\subsubsection{\textit{RQ$_2$} Why are these ecosystems dead?}

We have investigated what happened to the projects presented in RQ$_1$, to see if there is any continuation. 
In this regard, we investigate whether or not the original project is still alive, and if there have been any forks (i.e., others have taken the source code base and have evolved the software independently).
As shown in Table \ref{tab:RQ2}, new projects emerged in the aftermath of the dying ecosystem.

\begin{table}[t]
\begin{center}
\caption{Emergent Projects after the death of the ecosystem}
\begin{tabular}{l|c} \hline
System Name & \multicolumn{1}{c|}{Example Emergent Projects} \\ \hline \hline
Concurrent Versioning System (CVS) & CVSNT	\\ \hline
FireFoxOS & Panasonic variant, H5OS, KaiOS, Jio 	\\ \hline
Apache Geronimo & Tomcat, EJB, Derby 	\\ \hline
Maemo & MeeGo, Tizan, Mer 	\\ \hline
\end{tabular}    \label{tab:RQ2}
\end{center}
\end{table}

\paragraph{CVS}
Although the CVS project was discontinued, we find that due to the development of the Microsoft Windows, Linux, Solaris, HPUX, I5os and Mac OS X ports, evidence shows that CVS has split off into a separate project named CVSNT\footnote{website available at \url{https://www.march-hare.com/cvspro/} }, which is under current, active development (i.e., the latest update as of writing was April 2018).

\paragraph{FirefoxOS}
After the discontinuation of Firefox OS, several variants of the OS have emerged.
Panasonic will continue to develop the operating system for use in their Smart TVs, which runs My Home Screen, powered by the Firefox OS.
Acadine Technologies has derived their H5OS from Firefox OS as well. Li Gong, the founder of the company, has overseen the development of Firefox OS while serving as president of  the Mozilla Corporation. Alcatel OneTouch GO FLIP uses a fork called KaiOS\footnote{ website at \url{https://www.kaiostech.com/}}. In addition, in July 2017, it was reported that Indian telecom operator Jio would be launching new feature phone with OS derived from Firefox OS and the apps are purely in HTML5 and CSS.

\paragraph{Apache Geronimo}

The development of Apache Geronimo ceased around 2013, after its 3.0.1 release, when IBM and Oracle stopped to support the project in favor of their own technologies. Geronimo is not a single technology, but is the sum of many components, like Apache Tomcat\footnote{website as \url{http://tomcat.apache.org/}}, Apache EJB\footnote{website at \url{http://tomee.apache.org/tomcat-ejb.html}}, Apache Derby\footnote{website at \url{https://db.apache.org/derby/}}, among others. Many of these components are used in the implementation components of other frameworks as can be seen from http://arjan-tijms.omnifaces.org/2014/05/implementation-components-used-by.html.

\paragraph{Maemo}

In February 2010, the Maemo project from Nokia merged with Moblin to create the MeeGo mobile software platform under the umbrella of the Linux Foundation. However, the Maemo community continued to be active in Maemo. That is the reason why Nokia transferred the Maemo ownership first to the Hildon Foundation, and then to a German association called Maemo Community e.V. The last general assembly of this association has been in October 2017.

MeeGo\footnote{ A variant of MeeGo is Tizen \url{https://www.tizen.org/}} was cancelled in September 2011, although a community-driven successor called Mer\footnote{website as \url{http://www.merproject.org/}} was formed that. A Finnish start-up, Jolla, chose in 2013 Mer as the basis of the Sailfish OS operating system for their Jolla Phone smartphones. Another Mer derivative called Nemo Mobile is also currently developed actively.

\subsubsection{\textit{RQ$_3$} What can we learn from dead ecosystems?}

There is little to learn from dead ecosystems, because software ecosystems, at least those that are FLOSS, don’t die! In our quest for dead ecosystems, what we have found are that ecosystems that have been abandoned have evolved (if not completely, at least partially) with a given name. This means, that organizations and names are the ones that may disappear, but the technology can be found years later in other projects and developments. 
There are two main factors that may concur to explain this situation:

\begin{enumerate}
\item \textbf{Forks originating from the dead ecosystem.} The first one is the right to fork that exists (and is used) in FLOSS development. Although forking (i.e., splitting the community by taking the technology under a new name) is historically not welcome in the FLOSS community, it is understood in certain contexts. One of these situations is when the project is abandoned.
\item \textbf{Technological advancements.} The second one is related to the development of technologies. This requires time, much human labor and is maintenance intensive. A software is not only its development and its community. It is as well the number of tests and maturity that it has achieved. Successful FLOSS ecosystems have invested a large amount of effort in becoming mature. Even if its key players lose their interest in the technology and the community seems to shrink, there is always the source code, that is result of that effort. In addition, the investment in time and learning of other technologies results in inertia by those who are familiar with the technology. 
With ecosystems that have a large community, the probabilities of even a minor part of this community still interested in continuing with development is very high.
\end{enumerate}

\subsection{Conclusions}
FLOSS ecosystems are still too young to draw conclusions from our investigation, but as far as we have analyzed we have not found any (well-known) FLOSS ecosystem that can be considered dead (i.e., completely abandoned). For one or the other reason, the original software has evolved into other systems and communities and still serves, even if the importance of the project is not the one that used to be.

A lesson learned from our analysis is that if organizations want sustainability of a technology or application, they should strive for the ecosystem way. This is a lesson that could be of interest for consortia, public bodies, and companies wanting to set a standard. The network effects of developing a long-lasting software ecosystem is the probability that at least a small portion of the community keeps it alive. 
We have seen that this is the case from outdated technologies (like CVS) to hardware-linked software (such as Maemo). 

As there is a growing interest of corporations in FLOSS, such as the one that can be found in OpenStack, OW2, WebKit, among others, we are sure that the future will allow to have further examples of ecosystems and analyse how they evolve, even when their main promoters abandon.

\bibliography{sample}

\end{document}